\documentclass[5p,final,twocolumn]{elsarticle}
\usepackage{amsmath}
\usepackage{url}
\usepackage{listings}
\usepackage[english]{babel}

\journal{Computer Physics Communications}

\begin{document}

\begin{frontmatter}

\title{GPU-based simulation of the long-range Potts model \\ via parallel tempering}

\author{Attila Boer}
\ead{boera@unitbv.ro}
\address{``Transilvania'' University of Bra\c{s}ov, \\ Department of Electrical Engineering and Applied Physics, \\ B-dul Eroilor 29, 500036, Bra\c{s}ov, Romania}

\begin{abstract}
We discuss the efficiency of parallelization on graphical processing units (GPUs) for the simulation of the one dimensional Potts model with long-range interactions via parallel tempering. We investigate the behaviour of some thermodynamic properties, such as equilibrium energy and magnetization, critical temperatures as well as the separation between the first- and second-order regime. By implementing multispin coding techniques and an efficient parallelization of the interaction energy computation among threads, the GPU-accelerated approach reached speedup factors of up to 37.

\end{abstract}


\end{frontmatter}

\section{Introduction}\label{intro}

Increased computing capabilities over the last years have changed substantially the landscape of scientific computing. High performance graphical processing units (GPUs) have become an important part of computational clusters. High level programming languages, e.g. CUDA or OpenCL, have unlocked the computational power of GPUs and made it accessible to scientists.

Particularly the implementation of Monte Carlo simulations on GPUs can lead to increased performance due to the fact that the majority of these algorithms can be parallelized. Regrading spin systems the studies published in scientific journals were focused especially on systems with nearest neighbour interactions \cite{preis2009gpu,ferrero2012q}.

Macromolecules exhibiting long-range interactions have been simulated already on GPUs using molecular dynamics \cite{friedrichs2009accelerating} and the smooth particle mesh Ewald summation method \cite{harvey2009implementation}.
In the present paper we simulate the one dimensional Potts model with long-range interactions decaying as $1/r^{1+\sigma}$ using a parallel tempering algorithm implemented on GPUs. It is known from both classical \cite{wu1982potts} and nonextensive statistics \cite{boer2011monte} that the Potts model has a very rich thermodynamic behaviour.

The paper is structured as follows. In the first part we review some theoretical aspects regarding the Potts model and the parallel tempering algorithm. Section~\ref{sim} deals with the implementation details in CUDA. In Section~\ref{results} we present the principal results and benchmarks comparing the performance of the code with traditional CPU implementation. Finally in Section~\ref{concl} we draw some conclusions and final remarks.

The main result of this paper is the possibility of a good speedup on GPU for spin models with long-range interactions.
By exploiting the shared memory architecture of the GPU and using an efficient multispin coding scheme for the Metropolis step and the energy computation we were able to obtain a speedup factor of 37 compared with CPU implementations.

\section{The Model}\label{model}

The standard Potts model is a straightforward generalization of the well known Ising model. The one-dimensional $q$-state Potts model is a lattice where the spins $s_i$ can take the following values:
\begin{equation}
	s_i = 1, 2, \ldots, q
\end{equation}
The interaction Hamiltonian is given by the relation:
\begin{equation}
	\mathcal{H} = -\sum_{\substack{i,j\\ i\neq j}} J_{ij} \delta(s_i, s_j)
\end{equation}
where $J_{ij}$ is a coupling constant and $\delta(s_i, s_j)$ is the Kronecker delta:
\begin{equation}\label{eq:3}
	\delta(s_i,s_j)=
	\begin{cases}
		1 & \text{if } s_i=s_j     \\
		0 & \text{if } s_i\neq s_j 
	\end{cases}
\end{equation}
If we consider the long-range model, we must take into account the interactions between all spin pairs. The interaction is decaying with distance as $1/r^{1+\sigma}$ ($r$ being the distance between spins $i$ and $j$), so the coupling constant has the form 
\cite{glumac1998first}
\begin{equation}
	J_{ij} = \frac{1}{\left| i-j \right|^{1+\sigma}}
\end{equation}
For simulation purposes one usually resort to periodic boundary conditions. This can be approached by considering the interactions between all the spins in the original lattice of size $L$ and replacing the coupling constants 
$J_{ij}$ with
\begin{equation}
	J^* (r) = \sum_{n=-\infty}^{\infty} J(r+nL)
\end{equation}
This lead to the following expression for the coupling constants:
\begin{eqnarray}
	J^* (r) = \frac{1}{r^{1+\sigma}} &+& \frac{1}{L^{1+\sigma}} \left[
	\zeta\left( 1+\sigma, 1+\frac{r}{L} \right) \nonumber \right. \\
	&+& \left. \zeta\left( 1+\sigma, 1-\frac{r}{L} \right) \right]
\end{eqnarray}
where $\zeta(s,\alpha)$ is the Hurwitz zeta function 
\cite{reynal2004reexamination}.
\begin{equation}
	\zeta(s,\alpha) = \sum_{k=0}^\infty (k+\alpha)^{-s}
\end{equation}
The short-range Potts model undergoes a first order phase transition when $q>q_c$ and a second order phase transition when $q<q_c$, where the threshold value $q_c$ depends on the dimensionality of the lattice $d$ (i.e. $q_c=4$ for $d=2$). The long-range Potts model exhibits an interesting behavior: there is a so-called tricritical point at $\sigma_c$, depending on the value of $q$. For $\sigma < \sigma_c$ we have a first order transition, whereas for $\sigma > \sigma_c$ the transition is continuous.

\section{Parallel tempering (replica-exchange)}\label{pt}

Parallel tempering (a.k.a. replica exchange) is an algorithm which has been proposed to accelerate Monte Carlo simulations in systems with complex energy landscapes \cite{hukushima1996exchange,landau2005guide}. 

In this method we simulate in parallel multiple copies of the system, at different temperatures. For this stage we use the single spin-flip technique, based on the standard Metropolis algorithm, where the flips are accepted with probability
\begin{equation}
	p = \min \left( 1, e^{ - \beta \Delta E } \right)
\end{equation}
where $\Delta E$ represents the energy difference between the states. In the above relation $\beta=1/k_B T$ ($T$ denotes the temperature and $k_B$ stands for the Boltzmann constant).

After a certain number of Monte Carlo steps we attempt to exchange the states between neighboring temperatures, based on the Metropolis criterion with probability
\begin{equation}
	p = \min \left( 1, e^{ \left( E_i - E_j \right) \left( \beta_i - \beta_j
	\right) } \right)
\end{equation}
where $E_i$ and $E_j$ are the energies of the neighboring states, $\beta_i=1/k_B T_i$ and $\beta_j=1/k_B T_j$.

Parallel tempering is a powerful computational method, which is suitable for implementations on GPUs because different replicas can be run in parallel.

\section{GPU-based simulation}\label{sim}

\subsection{CUDA architecture}\label{cuda}
CUDA brings a new approach to parallel computing. It uses the Graphical Processing Unit (GPU) for scientific computing. The computing power of GPUs has increased rapidly in the last years and today they are faster than the CPU if we exploit their massively parallel nature.

From the software point of view CUDA is an extension of the C language, as well as a runtime library, which facilitates the general-purpose programming of NVIDIA GPUs. The parallel computation is organized using the abstractions of grids, blocks and threads. An entire grid is handled by a single GPU chip. Each multiprocessor on the GPU handles one or more blocks in a grid. Each multiprocessor is divided into a number of stream processors, each of them handling one or more threads in a block \cite{cudac-intro,cuda-intro}. 

Threads may only safely communicate with each other only if they are in the same thread block. The fastest communication between threads is done using shared memory. Shared memory is limited, usually 16 KB per multiprocessor on GT200-based chips or 48 KB on Fermi cards \cite{nvidia-fermi}. Often shared memory is simply not enough to store all the data that needs to be shared among the threads. In these cases we need to access global memory. Any thread in any block can read or write to any location in the global memory. Global memory is much larger than shared memory, however it is much slower.

\subsection{Implementation details}\label{impl}

For spin models with nearest neighbour interactions the checkerboard decomposition was implemented successfully on GPUs in the context of the Metropolis algorithm \cite{preis2009gpu,ferrero2012q} allowing considerable speedups compared with traditional CPU implementations. On the other hand for long-range models we must compute the interaction energy between all spin pairs in the lattice, so a checkerboard type decomposition is not suitable. In order to develop an efficient implementation we must seek other algorithms/computational techniques. For the simulations discussed in the present paper these techniques are:
\begin{itemize}
	\item[(i)] parallel tempering (well suited for parallel architectures)
	\item[(ii)] distribution of the energy computation among threads
	\item[(iii)] multispin coding.
\end{itemize}

Parallel tempering has been successfully implemented on GPUs for classical spin systems, such as the square-lattice Ising model with competing nearest-neighbor and next-nearest-neighbor interactions \cite{yin2009phase}, as well as the continuous Heisenberg and disordered spin-glass systems \cite{weigel2012performance}.

In this section we focus our attention on the implementation details in CUDA. All the computations on the GPU were performed using single-precision floating-point operations. The errors due to single floating point precision are smaller than $1.86 \cdot 10^{-3}\,\%$, which is acceptable. The standard deviation of rounding-off errors are negligible when compared to the error of the Monte Carlo simulation itself (under $7.83 \cdot 10^{-3}\,\%$).

Every thread block holds one replica. Usually the thread block size is smaller than the number of spins in the lattice, so every thread handles multiple spins.

Two different implementations were tested. In the first one spin values were stored in shared memory for the Metropolis steps and they were transferred back to global memory after $L$ Metropolis steps ($L$ being the number of spins in the lattice), which corresponds to an average of one Monte Carlo step per spin. In the second implementation all spin values were stored in global memory. The shared memory implementation proved to be much faster, as we will see in the benchmarking section.
The coupling constants were pre-calculated on the CPU using the Hurwitz zeta function from the GNU Scientific Library \cite{galassi2006gnu} and they were transferred from host memory to device global memory.

At the beginning of the Metropolis cycle we transfer the spin values from global to shared memory. Using bitwise operations we store multiple spin values in one \textsf{unsigned long long} type (see lines 3-12 in Listing~\ref{list:1}). At the end of the Metropolis cycle, which corresponds to an average of one Monte Carlo step per spin, we copy back the updated spin values to global memory.

\lstset{caption=Copying spin values from global to shared memory and store them back to global memory at the end of the Metropolis cycle,label=list:1}
\begin{lstlisting}
__shared__ unsigned long long sss[BLOCK_SIZE];
	...
// spt - number of spins per thread
int j = 0;
while (j < spt) {
	if (tid*spt+j < N) {
		spin = s[rid*N+tid*spt+j];
		spin = spin << j*3;
		if (j == 0) sss[tid] = spin;
		else sss[tid] = sss[tid] | spin;
	}
	j++;
}
	...
j = 0;
while (j < spt) {
	if (tid*spt+j < N) {
		int updated_spin = (sss[tid] >> j*3) & 7;
		s[rid*N+tid*spt+j] = updated_spin;
	}
	j++;
}
\end{lstlisting}

The interaction energy computation is parallelized among the threads in every block using a procedure similar with the one proposed by Gross et al. for the simulation of polymers \cite{gross2011massively}. 

At the first line from Listing~\ref{list:2} we allocate the shared memory. Every thread computes the interaction energy between a given number of spin pairs. Since multiple spin  values were stored in one \textsf{unsigned long long} type, we use bit shifting operations to extract the respective values for every spin in the lattice (see lines 16-17 in Listing~\ref{list:2}). Finally a parallel reduction is performed to obtain the total energy (not shown in the listing).
\lstset{stepnumber=2,caption=Parallelized energy computation,label=list:2}
\begin{lstlisting}
__shared__ float ee[BLOCK_SIZE];
	...
j = 0;
while (j < spt) {
	int spin = tid * spt + j;
	if (spin < N - 1) {
		for (int i = 0; i < N / 2; i++) {
			if (spin + 1 > i) {
				ind1 = i;
				ind2 = spin + 1;
			} else {
				ind1 = N - i - 1;
				ind2 = N - spin - 1;
			}
			int rij = ind2 - ind1;
			int spin1 = (s[ind1/spt] >> (ind1%spt)*3) & 7;
			int spin2 = (s[ind2/spt] >> (ind2%spt)*3) & 7;
			if (spin1 == spin2) ee[tid] -= JJJ[rij];
			/* JJJ[rij] are the coupling constants
			   stored in global memory */
		}
	}
	j++;
}
\end{lstlisting}

The Boltzmann acceptance ratio depends only on the change in energy, therefore in the Metropolis cycle we compute only the energies corresponding to the flipped spin before and after the spin change. At the end of the Metropolis cycle we calculate the total energy.

The replica exchange step is executed also on the GPU and consists of two stages. We try to exchange states between replicas $i$ and $i+1$ for which $i$ is even in the first stage, respectively $i$ is odd in the second stage (states were indexed from 0). These stages are executed using two consecutive kernels.

Random numbers are generated using the \textsf{CURAND} library \cite{curand}, which facilitates the simple and efficient generation of high-quality pseudorandom numbers. The resulting random numbers are stored in global memory. The default pseudorandom engine is an implementation of the Xorshift RNG \cite{marsaglia2003xorshift} and it produces higher quality random numbers than the Linear Congruential Generator.

The source code for all the simulations discussed in the present paper is available online \cite{lrpm-cuda}.

\section{Simulation results and benchmarks}\label{results}

\subsection{Thermodynamic properties}\label{thermo}

All the results discussed in the present section refer to the three-state Potts model ($q=3$).
Simulations were performed on lattice sizes ranging from 512 to 3584 and different values of $\sigma$ in the interval $[0.2,0.8]$. Every simulation involved $2\cdot 10^4$ thermalization steps and $5\cdot 10^4$ production steps. The number of replicas was set to $400$ with a uniform distribution of the temperatures. The following thermodynamic quantities were computed: equilibrium energy per spin $\langle E\rangle$, magnetization defined as \cite{reynal2004reexamination}
\begin{equation}\label{eq:10}
	M=\frac{q N_{\max}/N-1}{q-1}
\end{equation}
(where $N_{\max}=\max(N_1, N_2, \ldots, N_q)$, $N_i$ being the number of spins which has the value $i$), the fourth order cumulant of energy:
\begin{equation}\label{eq:11}
	U_E^{(4)}=\frac{\langle E^4 \rangle}{\langle E^2 \rangle^2}
\end{equation}
and the so called Binder cumulant:
\begin{equation}\label{eq:12}
	V_M^{(4)}=1-\frac{\langle M^4 \rangle}{3\langle M^2 \rangle^2}
\end{equation}
Errors were evaluated using the standard jackknife technique provided by the ALPS library (\url{http://alps.comp-phys.org/}) \cite{albuquerque2007alps,bauer2011alps}.

\begin{figure}[h]
	\centering
	\includegraphics[width=0.95\columnwidth,clip]{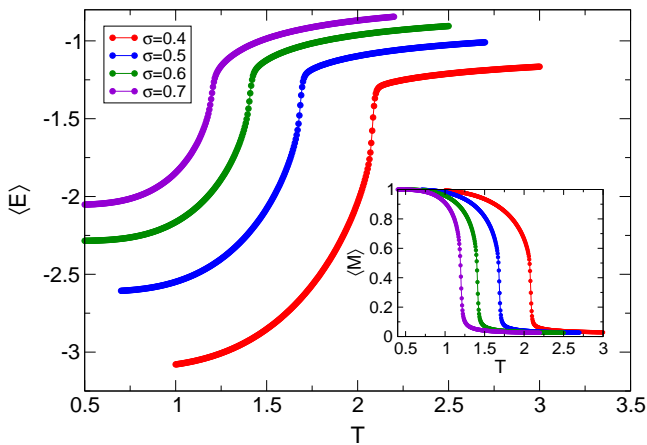}
	\caption{Mean value of energy per spin $\langle E\rangle$ and magnetization $\langle M\rangle$ (inset) versus temperature for $L=2048$ and different values of $\sigma$}
	\label{fig:1}
\end{figure}

Figure~\ref{fig:1} shows the temperature dependence of the equilibrium energy per spin and the magnetization for $L=2048$ and $\sigma=0.4,0.5,0.6,0.7$.

\begin{figure}[h]
	\centering
	\includegraphics[width=0.95\columnwidth,clip]{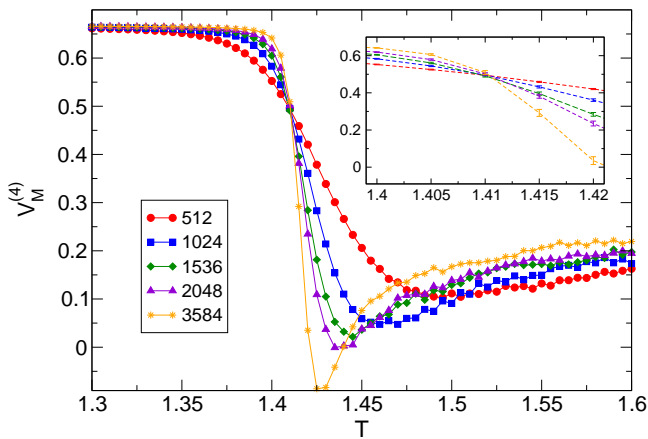}
	\caption{Binder cumulant vs. temperature for $\sigma=0.6$ and different lattice sizes}
	\label{fig:2}
\end{figure}
To evaluate the critical temperatures in the thermodynamic limit, we plotted the Binder cumulant as function of temperature for different lattice sizes (Fig.~\ref{fig:2}). The intersection point of the curves corresponds to the critical temperature. Values are presented in Table~\ref{tab:1}.

\begin{table}[h]
	\centering
	\begin{tabular}{p{1.5cm} p{1.5cm}}
		\hline\hline
		$\sigma$ & $T_c$        \\
		\hline
		$0.2$    & $3.955(5)$   \\
		$0.3$    & $2.7175(75)$ \\
		$0.4$    & $2.085(5)$   \\
		$0.5$    & $1.6875(75)$ \\
		$0.6$    & $1.4125(75)$ \\
		$0.7$    & $1.20(1)$    \\
		$0.8$    & $1.03(1)$    \\
		\hline\hline
	\end{tabular}
	\caption{Critical temperatures in the thermodynamic limit for different values of $\sigma$}
	\label{tab:1}
\end{table}

\begin{table}[h]
	\centering
	\begin{tabular}{p{1.5cm} p{2.2cm}}
		\hline\hline
		$\sigma$ & $U_{E}^{(4)}(L\to\infty)$ \\
		\hline
		$0.50$   & $1.022(1)$ \\
		$0.60$   & $1.0066(7)$ \\
		$0.65$   & $1.0021(5)$ \\
		$0.70$   & $1.0007(4)$ \\
		$0.72$   & $1.0006(3)$ \\
		$0.74$   & $1.0002(3)$ \\
        $0.76$   & $0.9999(2)$ \\
        $0.78$   & $0.9999(2)$ \\
		\hline\hline
	\end{tabular}
	\caption{Extrapolated values of the fourth order energy cumulant}
	\label{tab:2}
\end{table}

\begin{figure}[h]
	\centering
	\includegraphics[width=0.95\columnwidth,clip]{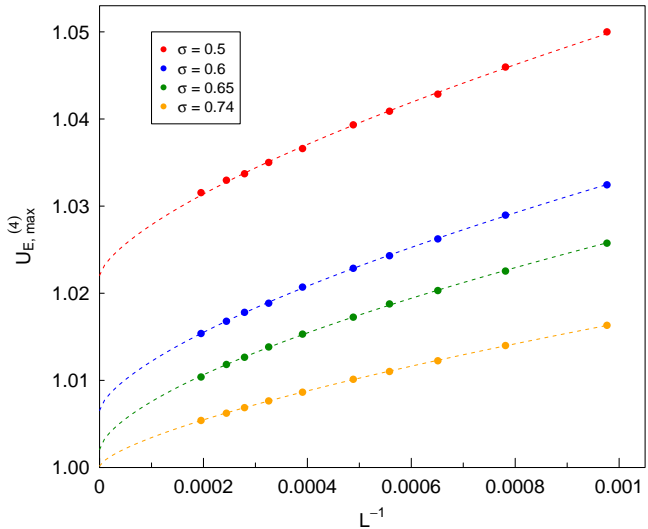}
	\caption{Fourth order energy cumulant versus inverse lattice size}
	\label{fig:3}
\end{figure}

To establish the order of the transition we studied the behaviour of
the fourth order cumulant of energy given by Eq.~\eqref{eq:11}. 
We considered 40 replicas around the pseudo-critical temperature and we performed simulations on lattice sizes ranging from 1024 to 5120 and different values of $\sigma$ in the interval $[0.5, 0.78]$. Every simulation involved $10^5$ thermalization steps and $2\cdot 10^5$ production steps. For each simulation we computed the maximum value of $U_E^{(4)}(L)$. If the transition is of first order the cumulant is different from $1$ in the thermodynamic limit, i.e. for $L\to\infty$. To establish the value of $U_{E}^{(4)}(L\to\infty)$ we fitted the curves corresponding to the dependence of $U_{E,\max}^{(4)}(L)$ versus $L^{-1}$ (Fig.~\ref{fig:3}) to the power law
\begin{equation}
	U_{E,\max}^{(4)}(L) = U_{E}^{(4)}(L\to\infty) + A \cdot L^{-B}
\end{equation}
$A$, $B$ and $ U_{E}^{(4)}(L\to\infty)$ being the fitting parameters.

Table~\ref{tab:2} summarizes the results obtained for $U_{E}^{(4)}(L\to\infty)$ and the corresponding errors. Curve fitting was performed using the Levenberg-Marquardt algorithm \cite{pressnumerical}. One can conclude that for $\sigma\leq 0.72$ the extrapolated value of the fourth order energy cumulant is different from 1 outside the estimated errors. 
For $\sigma \geq 0.74$ the extrapolated value has reached unity. Consequently the tricritical point which separates the first- and second-order regimes can be estimated as $\sigma_c=0.73(1)$. This result is in good agreement with the one obtained by Reynal et. al. using a multicanonical approach  \cite{reynal2004reexamination} and Glumac et. al. using short time dynamics \cite{uzelac2008short}.

\subsection{Benchmarks}\label{bench}

For benchmarking purposes short runs were performed on lattice sizes ranging from 512 to 10240, involving $10^4$ production steps for $L=512$ down to $50$ steps for $L=10240$. The number of thermalization steps was set to $20\%$ of the production steps, whereas the number of replicas was $400$ for all the simulations. 

We reserve three bits for every spin value, so the maximum number of spin values which can be stored in a $64$ bit integer is $21$ (if $q>7$ the number of bits should be increased). We must be careful to set a thread block size which is greater than $L/21$. Table~\ref{tab:3} shows the block sizes used for benchmarking.
\begin{table}[h]
	\centering
	\begin{tabular}{p{2.5cm} p{2.5cm}}
		\hline\hline
		$L$        & \textsf{Block size} \\
		\hline
		512-2560   & 128                 \\
		3072-5120  & 256                 \\
		5632-10240 & 512                 \\
		\hline\hline
	\end{tabular}
	\caption{Thread block sizes used in the benchmarking process}
	\label{tab:3}
\end{table}

\begin{table*}
	\centering
	\begin{tabular}{p{6cm} p{3.5cm} p{3.5cm}}
		\hline\hline
		                              & GPU                       & CPU                        \\
		\hline
		Name                          & NVIDIA Tesla C2050        & Intel Xeon X5660           \\
		Number of processors          & 14                        & 1                          \\
		Number of cores per processor & 32                        & 6 (only one used)          \\
		Clock speed                   & 1.15 GHz                  & 2.8 GHz                    \\
		RAM                           & 2687 MB                   & 24576 MB                   \\
		Max. threads per block        & 1024                      & -                          \\
		Shared memory per block       & 48 kB                     & -                          \\
		Compiler                      & NVCC 4.2 \footnotemark[1] & GCC 4.5.2 \footnotemark[2] \\
		\hline\hline
	\end{tabular}
	\caption{Characteristics of the computational setup}
	\label{tab:4}
\end{table*}
Table~\ref{tab:4} summarizes the characteristics of the computational setup used for the simulations.

\begin{figure}[h]
	\centering
	\includegraphics[width=0.95\columnwidth,clip]{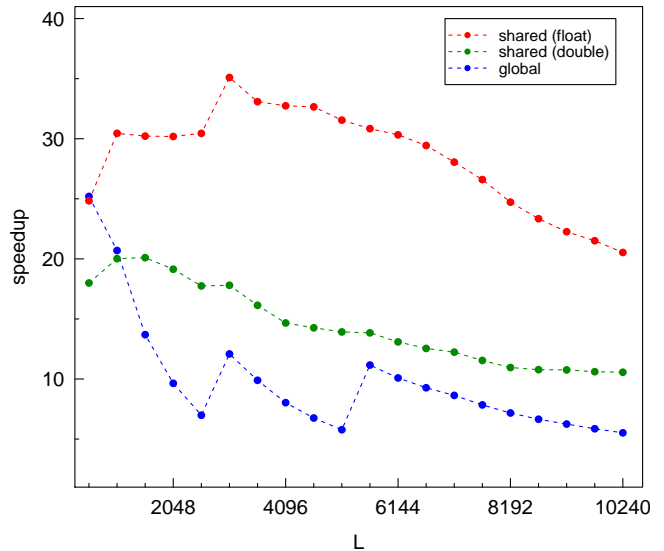}
	\caption{Speedup factor vs. lattice size}
	\label{fig:4}
\end{figure}

We have compared the performance of both the shared and global memory implementation with the corresponding CPU code. As we can see from the benchmarking results presented in Fig.~\ref{fig:4} the shared memory implementation with single floating point precision provides much better performance, the best speedup factor being $35.1$ for $L=3072$. For small lattice sizes ($L\leq 512$) the implementations perform roughly the same, but the performance of the global memory implementation decreases rapidly with the number of spins. For the shared memory implementation with double precision the speedup is approximately half of that corresponding to single floating point precision, but still outperforms the global memory variant. It is a known fact that GPUs provide much better performance if we use float instead of double. The errors due to floating point arithmetic are acceptable (see section~\ref{sim}), so we can choose the implementation which provides the best performance.

\footnotetext[1]{Compiler flags: \texttt{-O3 -arch=sm\_20}}
\footnotetext[2]{Compiler flags: \texttt{-O3 -ffast-math -funroll-loops}}

\begin{figure}[h]
  \centering
  \includegraphics[width=0.95\columnwidth,clip]{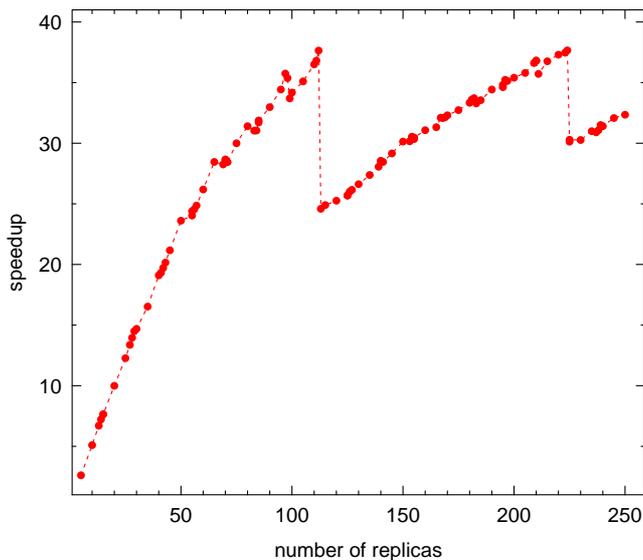}
  \caption{Speedup factor vs. number of replicas for $L=1024$ and $\sigma=0.2$}
  \label{fig:5}
\end{figure}

As can be expected, the speedup increases with the number of replicas (Fig.~\ref{fig:5}). The plot exhibits some maxima at certain values, which can be explained as follows. The Nvidia Tesla C2050 has 14 streaming multiprocessors (SMs). Every SM can access a number of 8 thread blocks simultaneously. Hence the maximum performance is achieved when the number of thread blocks is multiple of 112. The corresponding speedup factor is $37.6$.

\section{Conclusions and final remarks}\label{concl}

We performed Monte Carlo simulations of the one-dimensional Potts model with long-range interactions on graphical processing units using CUDA. We gain considerable speedups if we exploit the shared memory architecture of the GPU by implementing multispin coding techniques and an efficient parallelization of the interaction energy computation. 
From a thermodynamic point of view we evaluated the equilibrium energy and magnetization, respectively the critical temperature of transition for various values of $\sigma$. By studying the behaviour of the fourth order energy cumulant we established the threshold value of $\sigma$ which separates the first- and second-order regime.

One can conclude that graphical processing units can be very useful computing devices even for lattice spin models which exhibit long-range interactions. These systems can benefit also from the implementation of cluster algorithms on GPUs. This may be the subject of forthcoming studies.

\section*{Acknowledgments}
I would like to thank my wife, Milena, for her understanding and support during the writing of this paper.

I would like to thank the referee for his constructive criticism which helped to improve the manuscript.

\section*{References}
\bibliographystyle{model1-num-names}
\bibliography{references}

\end{document}